\documentclass[10pt,letterpaper]{article}
\usepackage[top=0.75in,left=0.75in,footskip=0.75in]{geometry}

\raggedright
\usepackage{amsmath,amssymb}

\usepackage{changepage}

\usepackage[utf8x]{inputenc}

\usepackage{textcomp,marvosym}

\usepackage{cite}

\usepackage{nameref,hyperref}

\usepackage[right]{lineno}

\usepackage{microtype}
\DisableLigatures[f]{encoding = *, family = * }

\usepackage[table]{xcolor}

\usepackage{array}

\newcolumntype{+}{!{\vrule width 2pt}}

\newlength\savedwidth

\usepackage[aboveskip=1pt,labelfont=bf,labelsep=period,justification=raggedright,singlelinecheck=off]{caption}

\makeatletter
\renewcommand{\@biblabel}[1]{\quad#1.}
\makeatother

\usepackage{lastpage,fancyhdr,graphicx}
\usepackage{epstopdf}
\fancyheadoffset[L]{2.25in}
\fancyfootoffset[L]{2.25in}
\usepackage{siunitx}

\begin{document}
\vspace*{0.2in}

\begin{flushleft}
{\Large
\textbf\newline{On the falsification of the pilot-wave interpretation of quantum mechanics and the meaning of the Born rule}
}
\newline
\\
Jakub M. Ratajczak
\\
\bigskip
Centre of New Technologies, University of Warsaw, Poland
\\
\bigskip
j.ratajczak@cent.uw.edu.pl, www.smearedgas.org

\end{flushleft}

\section{Abstract}
Quantum mechanics has lacked a widely recognized interpretation since its birth. Many of these are still under consideration because interpretations are tough or impossible to disprove experimentally. We show how to distinguish experimentally ones assuming tiny, localized particles from those postulating realistic, non-local wave functions. It is possible thanks to a recently developed model of optical transmittance of ultra-diluted gas. This model considers the influence of each gas particle wave function on transmitted light. Its quantitative predictions are not indifferent to the interpretation of quantum mechanics. We also refer to the results of a recent experiment founded on this theory. They are in line with the predictions, which suggests the correctness of the model. Such results, if confirmed, rule out interpretations of the type of pilot-wave. In the paper, we briefly explain both the experiment and the model it is founded on. We also discuss the meaning of the Born rule.

\section{Introduction}
Although quantum mechanics plays a fundamental role in modern physics, agrees with countless experiments, and is a foundation for many applications, it still lacks a widely recognized interpretation. Discussions on a proper one go back to the beginning of the quantum revolution. Currently, there are many rival interpretations, to name a few: Copenhagen \cite{sep-qm-copenhagen}, many worlds \cite{Everett1957}, pilot-wave \cite{Bohm1952}, objective collapse \cite{Ghirardi1986} \cite{Penrose1999} and many others, e.g., \cite{Cramer1986} \cite{Ballentine1970} \cite{Fuchs2014}. There are so many of them because they are difficult or impossible to disprove experimentally. This paper shows how to experimentally distinguish interpretations assuming tiny, localized particles from those postulating realistic, non-local wave functions. One of the most significant representatives of the latter is the pilot-wave interpretation. 

\subsection{Born rule}
First, we need to analyze the meaning of Born's rule, the rule that connects the mathematics of quantum mechanics to the outcome of an experiment. Born formulated the rule for any kind of quantum system and observable. So it allows us to regard a system consisting of only a single particle and consider its position, for simplicity, in one dimension:
\begin{equation}
P(x \in V, \Psi) = \int_{V}dx |\Psi(x)|^2~. \label{PEquation}
\end{equation}
In its original and still most popular understanding \cite{VonNeumann1955} \cite{landsman2009born} it means that if the system is in a state $\Psi$, then:

\begin{equation}
P(x \in V, \Psi)~is~the~probability~that~\underline{the~particle~is~found}~in~the~region~V.\label{BornRule}
\end{equation}

This formulation, with two different entities, namely the particle and the wave function, is required by the QM interpretations of the type of pilot-wave. According to \cite{Bacciagaluppi2012}  ``pilot-wave theories are no-collapse formulations of quantum mechanics that assign to the wave function the role of determining the evolution of (‘piloting’, ‘guiding’) the variables characterizing the system, say particle configurations''. 

Now let's try to dismiss ``particle'' and leave just $\Psi$ as the only real entity following Hobson \cite{Hobson2013}: ``$\Psi$ is a spatially extended field representing the probability amplitude (...) \emph{to interact at} $x$ rather than an amplitude for finding, upon measurement, a particle''. Then Eq.~(\ref{PEquation}) means that if the system is in a state $\Psi$, then:

\begin{equation}
P(x \in V, \Psi)~is~the~probability~that~\underline{it~interacts}~in~the~region~V. \label{HobsonRule}
\end{equation}

Here ``the particle'' means just some indivisible entity $\Psi$ refers to: the quantum. It doesn't refer to any spatial object like a tiny, ball-like thing. Note that the Born rule reformulated this way still preserves incredible, experimentally proven strength.

This way, we have two different formulations of the rule. They differ in the interpretation of what ``a particle'' and $\Psi$ are. Fig.~(\ref{fig:particlesChart}) illustrates it with two gas cloud diagrams. The first shows tiny particles guided by their wavefunctions, the ideal gas model. It follows the former interpretation of the Born rule, Ref.~(\ref{BornRule}). The second diagram follows the latter interpretation, Ref.~(\ref{HobsonRule}). For simplicity, but without losing generality, we assume the same spread for all particles. In the following section, we will refer to the optical depth rule \cite{Ratajczak2020b} \cite{Ratajczak2022} proposed for the gas with substantial wavefunctions' spread, the \emph{smeared gas}. This rule distinguishes both the gas models presented above. As a consequence, it will help us distinguish between both interpretations.

\section{Smeared gas} 
There is a proposed model \cite{Ratajczak2020b} \cite{Ratajczak2022} for the optical transmittance of ultra-diluted gas that considers the effects of each particle's wave function spreading. The spontaneous wavefunction spread results from solving the Schrödinger equation for a free particle. It applies to ultra-diluted gas because we recognize that the gas particles are independent of each other. 

This model predicts the dependence of the transmittance measurement on the size of the detector used.  Namely, a significant increase in the transmittance of such a gas is envisaged for a smaller detector. There are no such predictions from the classical models. It is valid in a nonrelativistic limit and for any type of gas. 

\begin{figure*}
\includegraphics[width=\textwidth]{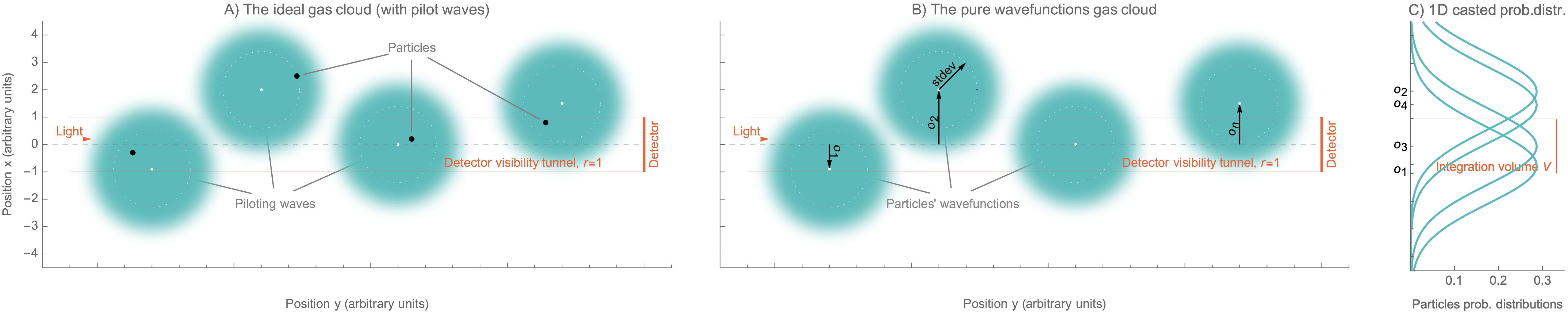}
\caption{The diagram shows a 2D gas cloud made of four independent particles. Their wavefunctions have the Gaussian distribution with uniform standard deviation $stdev$. Expectations are offset from the $x=0$ axis by $o_n$. The light direction, position of the light detector, and visibility tunnel (volume of integration $V$) are marked red. The detector is centered and placed parallel to the $X$ axis, far from the cloud to its right. Units are selected to have the detector diameter $r=1$. The detector is macroscopic, and the wavefunctions' spread is substantial: $stdev>r$. Diagram A) on the left shows the ideal gas model. Sample positions of ball-like particles are shown against the backdrop of their piloting wave functions. The middle chart B) does not suppose the existence of any finite-size entities. The right-hand part C) shows individual probability distributions cast along the $OY$ axis. Such a cast allows decreasing dimensionality of the problem by 1, here from 2D to 1D. It corresponds to Fig.~(3) in \cite{Ratajczak2022}.}\label{fig:particlesChart}
\end{figure*}

The model defines the optical transmittance $TR$ of a gas cloud as follows. This is \emph{the probability that a photon $\gamma$ coming from source that would have been detected by the detector in the absence of a cloud passes not scattered the entire $N$-element gas cloud and is detected by this detector}. Individual gas particles $A_n$ are independent, as are possible scattering events. This way, the transmittance may be assumed as a Markov chain of independent scatter-free events $(1-P_{A_n\gamma})$:

\begin{equation}
TR = \prod_{n=1}^N (1-P_{{A_n} \gamma})~, \label{TREquation}
\end{equation}

where $P_{A_n\gamma}$ is the probability of a photon (potentially measurable by the detector) being scattered by the $n$-th molecule of gas. According to the above transmittance definition, a scattering event referred by $P_{A_n\gamma}$ may occur only in some volume $V$, where both a (potentially detectable) photon and a particle have a chance for interaction, an interaction that influences measurement in the detector in any way. 

We will refer to the Born rule to unpack $P_{A_n\gamma}$:

\begin{equation}
P_{A_n\gamma}(t) = G \int_V |\Psi_{A_n}(\mathbf{r},t)|^2dr~, \label{P_Ag}
\end{equation}

where $V$ is a volume in which a detectable scattering event may occur. This volume is finite, and the diameter of the (macroscopic) detector constrains the boundaries of this volume. We call it \emph{the visibility tunnel}. Red lines in Fig.~(\ref{fig:particlesChart}) show this volume. A macroscopic detector allows us to neglect the higher-order corrections of nonclassic photon trajectories (path integrals). In other words, for considering $V$, we stay in the geometric optics approximation.

The $0<G<1$ coefficient adjusts the formula to real-world experimental conditions. It encodes all physical factors crucial for the specific setup, e.g., the photon wavelength and particles' cross-section. It is well known from spectrography that specific molecules ''like`` certain wavelengths more than others. It has nothing to do with wavefunctions' spread. It just reflects the strength of absorption lines. Note that this coefficient is assumed to be constant for a given setup. See \cite{Ratajczak2020b} and \cite{Ratajczak2022} for more.

Applying the well-known \cite{Shankar2011} solution of the Schrödinger equation for a free particle to Eq.~(\ref{P_Ag}), we obtain the lower limit of the transmittance of a gas cloud measured with a detector of any shape:

\begin{equation}
TR(\bar{t}) \ge \prod_{n=1}^N \left( 1-\frac{G}{4} \left[ erf \left( \frac{o_n-r}{\sqrt2 stdev_{A_n}(\bar{t})} \right) - erf \left( \frac{o_n+r}{\sqrt2 stdev_{A_n}(\bar{t})} \right) \right]^2 \right)~, \label{TRUpperLimit}
\end{equation}

where $r$ is half of the side of the smallest square circumscribed around the detector, $o_n$ is the $n$-th particle distance from the source-detector axis, $\bar{t}$ is the mean free time of the gas particles, and $stdev(\bar{t})$ is the expected standard deviation of the spread of a free particle wave packet $|\Psi|^2$. 

Note that the RHS of Eq.~(\ref{TRUpperLimit}) is the exact transmittance value for a square detector with a side of $2r$ and Gaussian distributions. The exact value of the transmittance for any shape of a detector or a wavefunction may be calculated from Eq.~(\ref{P_Ag}) using numerical integration.

The paper \cite{Ratajczak2020b} shows that the classic transmittance law, e.g., the Beer-Lambert law \cite{A.D.McNaught1997}, is just the first order approximation of the Eq.~(\ref{TREquation}) when a detector's diameter is much larger than $stdev(\bar{t})$. The model exhibits interesting properties when gas is thin enough, namely when $stdev(\bar{t})$ becomes comparable to (or larger than) the size of the detector. Then, transmittance depends on the size of the detector: the measured transmittance is larger, the smaller the detector is. The transmittance of such smeared gas becomes higher than expected by classic laws. It grows along with wavefunctions' spread. It may reach its limit up to 100\%, making such a gas practically undetectable with a detector of any reasonable finite size. None of these predictions are known by other transmittance models.

\subsection{Closed systems}
One may suspect that the effect of raising transmittance is merely caused by probability leaking out of the cloud. However, the paper \cite{Ratajczak2022}  (Eqs.~7-12) shows this is not the case. These unexpected transmittance properties also manifest themselves in a closed system. The closed system is a setup where the gas cloud is either closed in the box or very big. So big (say infinite) that one cannot neglect the probabilities of very distant (say infinite many) particles leaking into the visibility tunnel. In other words, the closed system is a system where the sum over $n$ of definite integrals over $V$ from probability distributions (Eq.~(\ref{P_Ag}) RHS) is constant.  

We found that in such a case, transmittance rises because the likelihood of a photon being absorbed (during the entire ``Markov process'' of passing by many wavefunctions) depends on \emph{the product} of the individual probabilities of absorption, while gas cloud mass depends on \emph{the sum} of masses. It is evident that the identical sum of two (equally long) sequences does not guarantee that both sequences' products are equal:
\begin{equation}
\left(\sum _{n=1}^{N} a_n=\sum _{n=1}^{N} b_n \right) \not \Rightarrow \left(\prod _{n=1}^{N} a_n=\prod _{n=1}^{N} b_n \right)~.
\end{equation}
Consequently, mass conservation (in the visibility tunnel) does not guarantee transmittance conservation. The product depends on how the individual distributions are divided. This division depends on (i) the shapes of the probability distributions and (ii) the width of the detector. 

\begin{figure*}
\includegraphics[width=\textwidth]{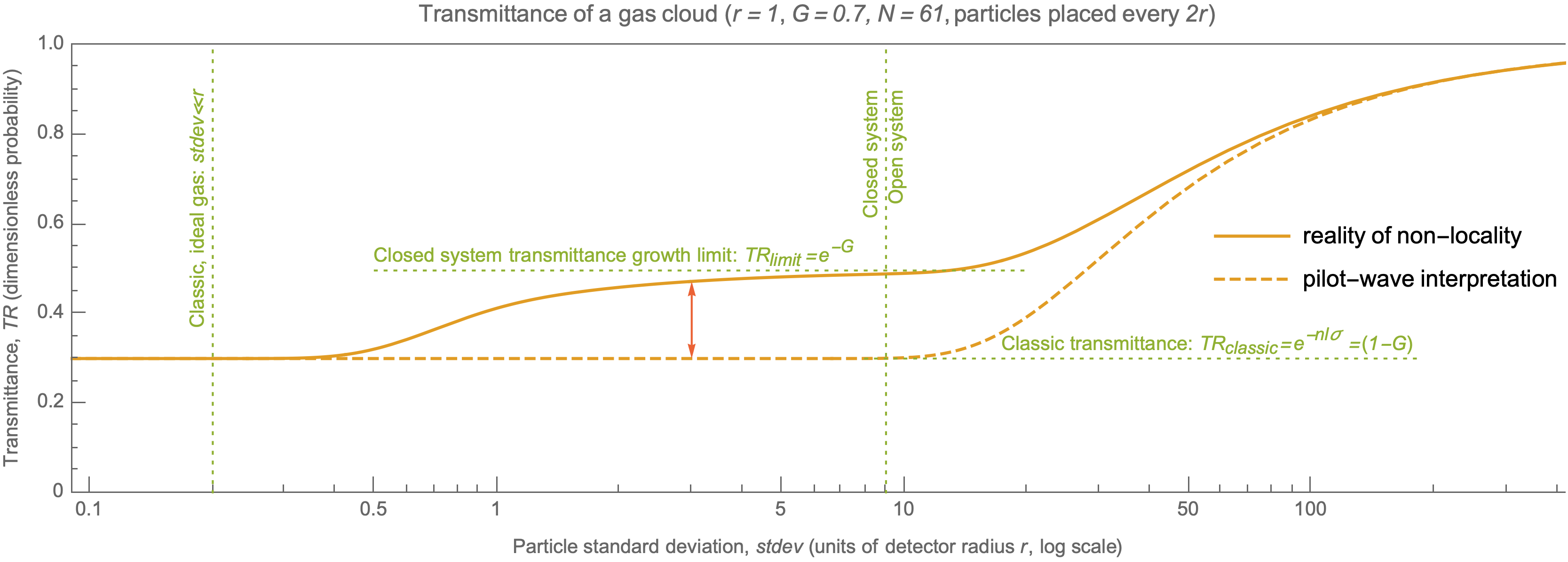}
\caption{The chart shows the dependence of transmittance on the gas particles' wavefunctions' spread. Spread, on the horizontal axis, is expressed as the Gaussian standard deviation. The detector has a fixed size. Its width is 2 ($r=1$), as the length unit is chosen to be equal to detector radius r. Thus the horizontal axis shows the wavefunction's spread to the detector's size ratio. The gas cloud is $N=61$ particles, and they are evenly spaced: $o_n-o_{n-1}=2$. The coefficient $G=0.7$ (after $TR_{classic}=30\%$). The particles have a 1D normal distribution where the standard deviation is also denominated in detector radius units ($r$). Different QM interpretations lead to different smeared gas transmittance model predictions. The solid line indicates the transmittance predicted with the assumption of ''non-locality reality``. The dashed line marks the transmittance assuming some small, ball-like objects are absorbing-or-not photons, i.e., according to the pilot-wave interpretation. The pilot-wave interpretation does not reveal any difference from classic, ideal gas transmittance for systems with mass conserved (see "Closed system" range). The red arrow points to a place where one can potentially conduct an experiment measuring transmittances that differentiates between the two interpretations. The same sample is presented in \cite{Ratajczak2022}.}\label{fig:transmittancesChart}
\end{figure*}

Direct evaluation of Eq.~(\ref{TRUpperLimit}) shows the direction of transmittance change with growing spread. It increases as Fig.~(\ref{fig:transmittancesChart}) presents. Moreover, we found \cite{Ratajczak2022} (Eq.~12) the limit of transmittance growth in a closed system:
\begin{equation}
TR_{limit}=\lim_{stdev \to \infty}TR_{closed}=e^{-G}=e^{(TR_{classic}-1)}~.\label{TRclosed}
\end{equation}

Although the transmittance in a closed system cannot increase up to 100\%, it can increase significantly (with growing spread) compared to the transmittance of a conserved mass ideal gas: $TR_{classic}<e^{(TR_{classic}-1)}$, because $0<TR<1$. This is the key to laboratory experiments trying to falsify the model. 

Note that we consider a closed system the really closed. For gas enclosed in a box, we neglect quantum tunneling by setting the enclosing box potential barrier high enough (macroscopical). For unbounded gas clouds, we assume the cloud diameter $D$ to be much bigger than a single particle spread: $D \gg stdev(\bar{t})$. 

This line of reasoning is valid only if we do not assume the existence of any finite-size balls ``merely guided'' by wave functions. Closed system transmittance is crucial for our discussion on discrimination of the interpretation, which we cover in the next section.

\section{Pilot-wave interpretation contradiction}
Fig.~(\ref{fig:particlesChart}) visualizes two different gas models depending on the adopted quantum mechanics interpretation. The left part (A) shows the piloting wave interpretation. It resembles the ideal gas model, even for significant wave functions' spreads. There are many tiny particles (black dots) in some random places.

Fig.~(\ref{fig:particlesChart}B) shows gas made of pure wave functions. They have a significant spread and are placed quite far from each other. It is not convenient in the literature to model gas this way because the gas in most applications is so dense that its particles do not experience much spread. For particles to gain such a spread, substantial mean free time is required. E.g., gas in deep space conditions or a dedicated laboratory setup. As shown in the previous section, we built the transmittance model upon the assumption of the reality of non-local wave functions. In particular, its consequence is the form of Eq.~(\ref{TRUpperLimit}). 

However, if the gas looked like Fig.~(\ref{fig:particlesChart}A), then the transmittance in a closed system would have to be equal to the classical one. According to the pilot wave interpretation, only the tiny, ball-like particles are responsible for any measurable interaction. In this case, it would be a mistake to rewrite the transmittance equation to the form Eq.~(11) in \cite{Ratajczak2022}. As a consequence, it would be improper to determine the transmittance limit of a closed system (for increasing spread) $TR_{limit}=\lim_{stdev \to \infty}TR_{closed}=e^{-G}$ as presented in Eq.~(\ref{TRclosed}) (see Eq.~(12) in \cite{Ratajczak2022}).  

There is another, simpler argument that transmittance can not change in a closed system. Note that ball-like particles can not escape such a system. Thus, as only they are responsible for scattering-or-not incoming photons, the good old Beer-Lambert law \cite{Bouguer1729} $TR_{classic}=e^{-\tau}=e^{\epsilon l c}$ enforces transmittance to stay classic because the concentration $c$ in the closed system stays constant.

Fig.~(\ref{fig:transmittancesChart}) plots expected transmittances for both interpretations on a single chart. They are presented as functions of particle spread $stdev(\bar{t})$. The left part of the chart, i.e., $stdev \lesssim 0.2$, shows transmittance for a classic, ideal gas. Gas particles are either not coherent, or their mean free time is too short. Both transmittances are the same and equal to the classic transmittance as predicted by the Beer-Lambert law \cite{A.D.McNaught1997}. The right part of the chart, i.e., $stdev \gtrsim 9$, shows both transmittances increase above $TR_{classic}$ and tend to 100\%. It is the open system case where there are no boundaries, and the spread starts to be significant. Gas particles stay coherent for a long time. Thus, the probability leaks out of the cloud. In turn, more leaks out of the tunnel of visibility than leaks in. On this particular chart, the open system starts with $stdev > 9$ just because we assumed the cloud is made of 61 particles and it extends symmetrically 30 particles on both sides of the visibility tunnel. Particles are evenly spaced 2 units of length apart from each other: $o_n-o_{n-1}=2$. Such a setup causes there is not enough probability (leaking into $V$) from distant particles when $stdev$ becomes greater than 9. If there were more particles even farther from the visibility tunnel $V$, then simply the spread would have to be greater for the probability to start leaking out of the cloud. However, note that $TR_{limit}$ would stay the same regardless of the cloud diameter. The plateau in the middle of the chart would be wider but not higher.

The middle part of Fig.~(\ref{fig:transmittancesChart}), where $0.2>stdev>9$, is the most interesting for distinguishing quantum mechanics interpretations. This is a closed system regime but with coherent gas particles. The straight dashed orange line presents transmittance according to the pilot wave interpretation. It equals the classic transmittance $TR_{classic}$: $TR_{closed}=TR_{classic}=(1-G)$. 

The solid orange line indicates the transmittance predicted with the assumption of ''non-locality reality``. It may be much higher than predicted classically, even in the closed system. The red arrow points to a potential transmittance experiment that differentiates between the two interpretations. As can be seen, the difference in transmittance can be substantial, reaching several dozen percentage points. Determining the transmittance of a gas is an easy, well-known experiment. Measuring the transmittance of a gas with a large spread-to-detector size ratio should determine which curve nature has chosen.

The value of the measured transmittance equal to the classical transmittance would mean that the understanding presented in our papers \cite{Ratajczak2020b} and \cite{Ratajczak2022} is wrong and Born's rule is about finding a particle as in Ref.~\ref{BornRule}. A higher value of the measured transmittance would indicate the correctness of the proposed model and the understanding of $P$ as in Ref.~\ref{HobsonRule}. It is possible to distinguish both cases experimentally.

Achieving a sufficiently high ratio of the wave function spread to the diameter of the detector in a modern laboratory is not a big challenge. In the experiment \cite{Ratajczak2021} we tested the transmittance dependence on the detector size for a water vapor NIR absorption line $\lambda=\SI{1.36}{\micro\metre}$. We measured in parallel transmittances using a pair of detectors with different diameters ranging from \SI{2}{\micro\metre} to \SI{200}{\micro\metre}. The anticipated mean free path was $\sim$\SI{37}{\centi\metre} thanks to low enough $\sim$\SI{e-2}{\milli\bar} vacuum and large enough $\sim$\SI{1}{\cubic\metre} vacuum chamber. Consequently, we achieved a significant spread $stdev(\bar{t})\sim\SI{14}{\micro\metre}$. 

This experiment showed that transmittance in the closed system is higher than expected with classical predictions with $>5\sigma$ statistical significance. Apart from a possible flaw in the experiment, we know no other model that predicts the phenomena observed.  It indicates that the solid line in Fig.~(\ref{fig:transmittancesChart}) correctly describes the transmittance of the system. We treat it as a promising sign. As far as we know, this was the first experiment of its kind. Further experimentation to confirm our findings is undoubtedly necessary. If confirmed it would both verify the proposed transmittance model and exclude the pilot-wave interpretation.

\section{Summary}
In this paper, we showed that it is possible to design an experiment to distinguish the pilot-wave interpretation of quantum mechanics from an interpretation that does not assume the existence of any tiny, local entities (ball-like particles). For this purpose, we employed the newly developed transmittance model of ultra-diluted gases \cite{Ratajczak2022}. This model considers the quantum spread of individual gas particles assuming their coherent behavior. The model allows differentiating if there are or there are not present ball-like particles in a gas cloud. The model is falsifiable and it is possible to design an experiment. We showed how to interpret the results of such an experiment to distinguish both interpretations.

Moreover, we referred to the actual experiment \cite{Ratajczak2021} conducted according to the abovementioned model. To the best of our knowledge, this is a rare example of an experiment not indifferent to the interpretation of quantum mechanics. Results of the experiment with high statistical significance follow predictions of the model and thus contradictt the pilot-wave interpretation of quantum mechanics.

\bibliography{library.bib}

\end{document}